\documentclass{article}

\usepackage[dvips]{graphics}
\usepackage{graphicx}
\usepackage[dvips]{color}

\catcode`\@=11

\@addtoreset{equation}{section} \catcode`\@=10

\def\half{\mbox{$\frac{1}{2}$}}
\def\d{\mbox{\rm d}}

\def\dddot#1{\mathinner{\buildrel\vbox{\kern5pt\hbox{...}}\over{#1}}}
\def\ddddot#1{\mathinner{\buildrel\vbox{\kern5pt\hbox{....}}\over{#1}}}

\begin{document}

\begin{center}
{\Large An algebraic approach to laying a ghost to rest}\\[3 mm ]
{\large MC Nucci${} ^ {\dagger} $ and PGL Leach${} ^ {\ddagger} $\footnote {permanent address: School of Mathematical Sciences, University of KwaZulu-Natal,
Private Bag X54001 Durban 4000, Republic of South Africa}}\\[2 mm ]
{\large ${} ^ {\dagger} $ Dipartimento di Matematica e Informatica, Universit\`a di Perugia, 06123 Perugia, Italy\\Email: nucci@unipg.it}\\[2 mm ]
{\large ${} ^ {\ddagger} $ DICSE, University of the Aegean, Karlovassi 83 200, Greece\\Email: leachp@ukzn.ac.za; leach@math.aegean.gr}
\end{center}

\textbf{{Abstract:}} In the recent literature there has been a
resurgence of interest in the fourth-order field-theoretic model
of Pais-Uhlenbeck \cite {Pais-Uhlenbeck 50 a}, which has not had a
good reception over the last half century due to the existence of
{\em ghosts} in the properties of the quantum mechanical solution.
Bender and Mannheim \cite{Bender 08 a} were successful in
persuading the corresponding quantum operator to `give up the
ghost'. Their success had the advantage of making the model of
Pais-Uhlenbeck acceptable to the physical community and in the
process added further credit to the cause of advancement of the
use of ${\cal PT} $ symmetry. We present a case for the acceptance
of the Pais-Uhlenbeck model in the context of Dirac's theory by
providing an Hamiltonian which is not quantum mechanically
haunted. The essential point is the manner in which a fourth-order
equation is rendered into a system of second-order equations. We
show by means of the method of reduction of order \cite {Nucci}
that it is possible to construct an Hamiltonian which gives rise
to a satisfactory quantal description without having to abandon
Dirac.
 \strut\hfill

\textbf {{Keywords:}}  Quantisation; Lie symmetry; ${\cal
PT}$-symmetry

\strut\hfill

\section{Introduction}

In one of the great books of Science Dirac \cite {Dirac 32 a}
provided a theoretical environment for the rapidly developing
subject of Quantum Mechanics based upon an interpretation of
Mechanics elaborated by Hamilton \cite {Hamilton} almost a century
earlier in terms of operators.  In the process and following the
audacious line of thinking established by his predecessor at
Bristol, Oliver Heaviside, Dirac introduced the distributions
which were the bane of mathematicians until Laurent Schwartz \cite
{Schwartz 50 a, Schwartz 51 a} provided a theoretical
justification and George Temple \cite {Temple} clarified that
justification.  In his monograph Dirac took the classical
Hamiltonian as the energy and it is by no means obvious that he
ever considered the possibility of any other Hamiltonian function
as the basis for his operators.  Moreover one should emphasise
that the energy was conserved so that the approach of Bateman
\cite {Bateman 31 a} to the quantisation of the damped linear
oscillator could scarcely be regarded as being within the purview
of Dirac's theory\footnote {Bateman's contribution seems already
to have become lost by the following decade since the names often
associated with this problem are Caldriola \cite {Caldriola 41 a}
and Kanai \cite {Kanai 49 a}.  The question of how to deal
properly with dissipation in Quantum Mechanics became a lively
issue in the following decades and is an interesting study in
itself, but it is not immediately germane to our present topic.}.

\strut\hfill

There arose models of quantum mechanical systems which did not
coincide with Dirac's {\it obiter dictum} of a Hamiltonian as the
energy.  Indeed the direct association of a Hamiltonian with such
a model was impossible.  A well-known example is the model of
Pais-Uhlenbeck \cite {Pais-Uhlenbeck 50 a} in which the Action is
given by
\begin {equation}
A = \half\gamma\int\left\{\ddot {z} ^ 2 - \left (\Omega_1 ^ 2+\Omega_2 ^ 2\right)\dot {z} ^ 2
+ \Omega_1 ^ 2\Omega_2 ^ 2z ^ 2\right\}\d t \label {1.1}
\end {equation}
so the Euler-Lagrange equation is of the fourth order, being
\begin {equation}
\ddddot {z} + \left (\Omega_1 ^ 2+\Omega_2 ^ 2\right)\ddot {z} + \Omega_1 ^ 2\Omega_2 ^ 2z = 0. \label {1.2}
\end {equation}
To bring this model within the context of Hamiltonian Theory Mannheim and Davidson \cite {Mannheim 05 a} and Mannheim \cite {Mannheim 07 a} introduced a new variable, $y $, to describe the model in terms of a system with two degrees of freedom.  The resulting Hamiltonian is \cite {Bender 08 a}
\begin {equation}
H = \frac {p_y ^ 2} {2\gamma} + p_zy + \frac{\gamma}{2}\left
(\Omega_1 ^ 2+ \Omega_2 ^ 2\right)y ^ 2 - \frac{\gamma}{2}\Omega_1
^ 2\Omega_2 ^ 2z ^ 2. \label {1.3}
\end {equation}

\strut\hfill

The problem with the model of Pais-Uhlenbeck is that it possesses
`ghost' states, {\it ie} the norm of the (quantum) state is
negative.  This is not acceptable mathematically as well as
physically since the whole concept of a norm is rooted in the
essence of being nonnegative.  Consequently the model has been
regarded as unphysical.  Even in the representation as a
two-degree-of-freedom system, (\ref {1.3}), there are problems.
According to Bender and Mannheim \cite {Bender 08 a} there are two
possibilities depending upon the operators which annihilate the
groundstate.  The first gives a negative norm\footnote {We
maintain the definition of norm in the sense of Dirac.}.  The
second avoids that problem by giving a spectrum which is unbounded
from below.  Neither option could be described as acceptable!

\strut\hfill

In the last decade or so there has been a considerable expansion of what has been termed
 ${\cal PT} $ quantum mechanics.
 This is not the place to enter into a disquisition on the subject.
 We refer the reader to the references given in Bender and Mannheim \cite {Bender 08 a}.
It suffices to observe that in ${\cal PT} $ quantum mechanics the
concept of an hermitian operator in the sense of Dirac\footnote
{This is the perfectly normal one in Mathematics of being
invariant under the two processes of transposition of the matrix
and the taking of the complex conjugate of the elements.} is
replaced by definitions based more in Physics.  The ${\cal P}$
operator is that of reflection in space and ${\cal T}$ is the
operation of time reversal.

\strut\hfill

The case made by Bender and Mannheim \cite {Bender 08 a} is that,
if one interprets the Pais-Uhlenbeck model in terms of ${\cal PT}
$ quantum mechanics, the problem of the existence of ghost states
is removed.  Thereby the validity of the model is enhanced.

\strut\hfill

In this Letter we do not argue against the use of ${\cal PT} $
quantum mechanics for it does seem to be able to explain many
phenomena which experience difficulties when discussed in terms of
the model of Dirac.  However, we present a case for the acceptance
of the Pais-Uhlenbeck model in the context of Dirac's theory.  The
essential point is the manner in which the fourth-order equation,
(\ref {1.2}), is rendered into a system of second-order equations.
We show by means of the method of reduction of order \cite {Nucci}
that it is possible to construct an Hamiltonian which gives rise
to a satisfactory quantal description without having to abandon
Dirac.

\strut\hfill

The plan of this Letter is simple. In the next Section we
demonstrate the construction of a suitable Lagrangian by means of
the technique of the method of reduction of order.  The
Hamiltonian follows in the natural way.  In the following Section
we briefly perform the obvious quantisation procedure in the
manner of Dirac to demonstrate explicitly that we have a
physically consistent description in the case of the
Pais-Uhlenbeck model.

\strut\hfill

Before we commence it is appropriate that we remind the reader of
the importance of symmetry in the analysis of differential
equations, be they from classical or quantal origins.  When one is
dealing with Lagrangian, hence Hamiltonian, systems the importance
of symmetry seems to be enhanced.  One can make many Lagrangians
for a given system \cite {Nucci again again}.  They need not be
identical in their properties in terms their symmetries.  It is a
present matter for investigation of the implications of a
differing number of Noether symmetries for Lagrangians describing
the same physical system.

\section{Lagrangian description for the Pais-Uhlenbeck model}

The method of reduction of order has been used in a number of
papers in recent years (see \cite {Nucci again} and references
within) and we briefly summarise the method.  Further details may
be found in the papers cited.  Given a system of ordinary
differential equations of greater than the first order (at least
one of the system), the system is replaced by another system of
equations of the first order by the introduction of the requisite
number of new dependent variables.  At least one of the variables
is removed so that there is at least one equation of the second
order.  This enables one to make sensible use of Lie's theory of
continuous groups since the number of point symmetries is then
finite.  Armed with the symmetries one can then make a further
analysis by means of the standard theory.  In the present case we
are concerned with the transition from a classical description to
a quantum mechanical description.  Obviously symmetry plays an
important role in both descriptions.  Our initial problem in terms
of the method of reduction of order of the fourth-order equation,
(\ref {1.2}), is to find a pair of second-order equations for
which an obtainable Lagrangian description exists.  Then we have
the question of the determination of an Hamiltonian and is
quantisation to an operator which has sensible properties.

\strut\hfill

We observe that generally speaking there are many possible ways in
which a system of first-order equations can be constructed from
the original fourth-order equation and then retreaded as a pair of
second-order equations.  The way\footnote {Evidently without an
appreciation of the general principles of the method of reduction
of order.} chosen by Mannheim and Davidson \cite {Mannheim 05 a}
and Mannheim \cite {Mannheim 07 a} was demonstrated by Bender and
Mannheim \cite {Bender 08 a} to be unsatisfactory in terms of the
prescriptions of Dirac\footnote {Although it is not stated so in
\cite {Bender 08 a}, in \cite {Mannheim 05 a} and \cite {Mannheim
07 a} we are informed that the Hamiltonian description is obtained
using the method of Ostrogradsky, as described in the classical
text by Whittaker \cite {Whittaker 44 a}.}.  Our systematic
approach through the method of reduction of order demonstrates a
more than somewhat different result.

\strut\hfill

We introduce new variables to render (\ref {1.2}) as a system of four first-order equations.
 To maintain a consistent notation we write $z = w_1 $ and continue as
\begin {eqnarray}
& &\dot {w}_1 = w_2 \label {2.1} \\
& &\dot {w}_2 = w_3 \label {2.2} \\
& &\dot {w}_3 = w_4 \label {2.3} \\
& &\dot {w}_4 = -\left (\Omega_1 ^ 2+\Omega_2 ^ 2\right) w_3 - \Omega_1 ^ 2\Omega_2 ^ 2w_1.  \label {2.4}
\end {eqnarray}
In the usual application of the method of reduction of order an ignorable coordinate can the eliminated, but in this instance we do not wish to remove $t $ as the independent variable.  As our intermediate aim is to construct an Hamiltonian, we wish to rewrite the system of four first-order equations as a pair of second-order equations as a prelude to the construction of a first-order Lagrangian in two dependent variables.  If we eliminate $w_2 $ and $w_4 $, the system, (\ref {2.1}-\ref {2.4}), becomes
\begin {eqnarray}
& & \ddot {w}_1 = w_3 \label {2.5} \\
& & \ddot {w}_3 = -\left (\Omega_1 ^ 2+\Omega_2 ^ 2\right)w_3 - \Omega_1 ^ 2\Omega_2 ^ 2w_1. \label {2.6}
\end {eqnarray}
Continuing in the spirit of the method of reduction of order
 we introduce two more suitable dependent variables by means of
\begin {eqnarray}
& & w_1 = r_1 - r_2 \label {2.7} \\
& & w_3 = -\Omega_1 ^ 2r_1+ \Omega_2 ^ 2r_2, \label {2.8}
\end {eqnarray}
from which it is evident that we are dealing with the case
$\Omega_1 > \Omega_2 $. Equations (\ref {2.7}) and (\ref {2.8})
are written in terms of the new variables and rearranged to give
\begin {eqnarray}
& & \ddot {r}_1 = -\Omega_1 ^ 2r_1 \label {2.9} \\
& & \ddot {r}_2 = -\Omega_2 ^ 2r_2, \label {2.10}
\end {eqnarray}
which obviously describes a two-dimensional anisotropic oscillator.

\strut\hfill

A first-order Lagrangian for system (\ref {2.9},\ref {2.10}) is
\begin {equation}
L = \half\left (\dot {r}_1 ^ 2+ \dot {r}_2 ^ 2 - \Omega_1 ^ 2r_1 ^ 2 - \Omega_2 ^ 2r_2 ^ 2\right) \label {2.11}
\end {equation}
from which it is obvious that the canonical momenta are
\begin {equation}
p_1 = \dot {r}_1\quad \mbox {\rm and} \quad p_2 = \dot {r}_2 \label {2.12}
\end {equation}
and so the Hamiltonian is
\begin {equation}
H = \half\left (p_1 ^ 2+p_2 ^ 2+ \Omega_1 ^ 2r_1 ^ 2+ \Omega_2 ^ 2r_2 ^ 2\right). \label {2.13}
\end {equation}
We note that the Hamiltonian of (\ref {2.13}) belongs to the class of classical Hamiltonians envisaged by Dirac.

\strut\hfill

\section {Quantum Mechanical Formulation}

Since the Hamiltonian, (\ref {2.13}), is under the \ae gis of
Dirac's canon, we may apply the standard methods of quantisation
to obtain the Schr\"odinger equation
\begin {equation}
2i\frac {\partial u} {\partial t} = -\frac {\partial ^ 2u} {\partial r_1 ^ 2} -\frac {\partial ^ 2u} {\partial r_2 ^ 2} + \left (\Omega_1 ^ 2r_1 ^ 2+ \Omega_2 ^ 2r  ^ 2\right)u. \label {3.1}
\end {equation}
This has seven Lie point symmetries\footnote {Five of these
correspond to the Noether point symmetries of the Lagrangian,
(\ref {2.11}), as one would expect \cite {Lemmer 99 a}.  The
remaining two are generic to linear evolution equations.  The
corresponding system, (\ref {2.9},\ref {2.10}), possesses seven
Lie point symmetries \cite {Gorringe 88 a}.  Five of these are the
same as the Noether point symmetries.  The additional two are a
consequence of the linearity of the equations in the dependent
variables.}.  They are
\begin {eqnarray}
& & \Gamma_{\pm 1} = \exp [\pm i\Omega_1t ]\left\{\mp \partial_{r_1} + \Omega_1r_1u\partial_u\right\} \label {3.2} \\
& & \Gamma_{\pm 2} = \exp [\pm i\Omega_2t ]\left\{\mp \partial_{r_2} + \Omega_2r_2u\partial_u\right\} \label {3.3} \\
& & \Gamma_3 = i\partial_t \label {3.4} \\
& & \Gamma_4 = u\partial_u \label {3.5} \\
& & \Gamma_5 = f (t,r_1,r_2)\partial_u, \label {3.6}
\end {eqnarray}
where $f $ is a solution of (\ref {3.1})\footnote {Naturally equation (\ref {3.1}) comes with boundary conditions.  The symmetry $\Gamma_5 $ is not confined to being in terms of functions which satisfied the boundary conditions.}.

\strut\hfill

Since the Hamiltonian, (\ref {2.13}), is separable in the variables which we have selected, the determination of Dirac's creation and annihilation operators follows immediately from the symmetries (\ref {3.2}) and (\ref {3.3}) which are their progenitors \cite {Lemmer 99 a, Andriopoulos 05 a}.  The energy follows as the eigenvalue of the action of $\Gamma_3 $ on a solution.  The groundstate, as an explicitly time-dependent function, is obtained using $\Gamma_1 $ and $\Gamma_2 $ as follows.

\strut\hfill

The invariants of $\Gamma_1 $ are determined by the solution of the associated Lagrange's system
\begin {equation}
\frac {\d t} {0} = \frac {\d r_1} {-1} = \frac {\d r_2} {0} = \frac {\d u} {\Omega_1r_1u}  \label {3.7}
\end {equation}
and are $t $, $r_2 $ and $v = u\exp \left [\half\Omega_1r_1 ^ 2\right ] $.  In terms of these invariants $\Gamma_2 $ is \\$\exp [i\Omega_2t ]\left (-\partial_{r_2} + \Omega_2r_2v\partial_v\right) $.  The solution of the corresponding associated Lagrange's system gives the invariants $t $ and $w = v\exp\left [\half\Omega_2r_2 ^ 2\right ] $.  The double reduction of (\ref {3.1}) to an ordinary differential equation is achieved by writing
\begin {equation}
u = \exp\left [ -\half\left (\Omega_1r_1+ \Omega_2r_2 ^ 2\right)\right ] f (t). \label {3.8}
\end {equation}
The reduced equation is
\begin {equation}
\frac {\dot {f}} {f} = \frac {1} { 2i}\left (\Omega_1+ \Omega_2\right), \label {3.9}
\end {equation}
where the overdot denotes differentiation with respect to $t $, and has the solution $f (t) = \exp\left [ -\half\left (\Omega_1+ \Omega_2\right) it\right ] $.

\strut\hfill

The groundstate wavefunction is, up to the normalisation factor,
\begin {equation}
u_0 = \exp\left [-\half\left (\Omega_1+ \Omega_2\right) it -\half\left (\Omega_1r_1+ \Omega_2r_2 ^ 2\right)\right ]. \label {3.10}
\end {equation}
The action of $\Gamma_3 $ gives the groundstate energy through
\begin {eqnarray}
& & \Gamma_3u_0 = \half\left (\Omega_1+ \Omega_2\right)u_0 \label {3.11} \\
\Longrightarrow & & \phantom{\Gamma} E_0 = \half\left (\Omega_1+ \Omega_2\right). \label {3.12}
\end {eqnarray}

\strut\hfill

Further eigenstates are obtained by the actions of $\Gamma_{- 1} $ and $\Gamma_{-2} $ which act as creation operators for the time-dependent Schr\"odinger equation, (\ref {3.1}).  Since these are simply combinations of the eigenstates of two harmonic oscillators, there are no difficulties with the positivity of the energy spectrum, which is obtained by the action of $\Gamma_3 $ on the created eigenstates, and the nonnegativity of the norm of the wavefunction in the sense of Dirac.

\strut\hfill

\section {Observations}

It has not been our intention to belittle the value of the use of
${\cal PT} $  symmetry in the resolution of some questionable
problems in quantum mechanics.  What we have specifically desired
to demonstrate is that the model of Pais-Uhlenbeck can be rendered
into an Hamiltonian form which can be quantised and lead to
results after a mathematical analysis which are consistent with
the physical principles underlying the model.  In this respect we
are guided by the principle of the maintenance of symmetry in
going from the classical model to the corresponding Schr\"odinger
equation.  The second-order Lagrangian for the Action Integral in
(\ref{1.1}) possesses five Noether point symmetries.  After the
application of the method of reduction of order to obtain the
Lagrangian (\ref {2.11}) we find that this Lagrangian possesses
seven Noether point symmetries.  This increase is not to be
unexpected since we have essentially introduced generalised
symmetries\footnote {The application of the method of Ostrogradsky
does the same in that the second dependent variable introduced is
$\dot {z} $.  This does not mean that the construction of the two
first-order Lagrangians are equivalent with respect to point
symmetries.}.  In our approach these Noether symmetries are
preserved in the transition to quantum mechanics as the nongeneric
symmetries of the Schr\"odinger equation corresponding to {\it
our} Hamiltonian.  It is not at all obvious that this be the case
for the Hamiltonian used in \cite {Mannheim 05 a}, \cite {Mannheim
07 a} and \cite {Bender 08 a}.

\strut\hfill

We would propose as a general principle that any technique of
quantisation from a corresponding classical system be such that
there is preservation of the Noether point symmetries of the
classical Lagrangian in the nongeneric Lie point symmetries of the
corresponding Schr\"odinger equation.  It is unfortunate that at
this stage we cannot remove the `would' from that which we propose
since it is a very general question, but we are working on it.

\strut\hfill

\section*{Acknowledgements}

PGLL thanks DICSE, University of the Aegean, for the provision of facilities whilst this work was undertaken.  The University of KwaZulu-Natal and the National Research Foundation of South Africa are thanked for their support.

\strut\hfill


\begin {thebibliography} {99}

\bibitem{Andriopoulos 05 a}
Andriopoulos K \& Leach PGL (2005) Wave-functions for the time-dependent linear oscillator and Lie point symmetries {\it Journal of Physics A: Mathematical and General} {\bf 38} 4365-4374

\bibitem {Bateman 31 a}
Bateman H (1931) On dissipative systems and related variational principles {\it Physical Review} {\bf 38} 815-819

\bibitem {Bender 08 a}
Bender CM \& Mannheim PD (2008) Giving up the ghost {\it Journal of Physics A: Mathematical and Theoretical} {\bf 41} 304018 (7pp)

\bibitem {Caldriola 41 a}
Caldriola P (1941) Forze non conservative nella meccanica quantistica {\it Il Nuovo Cimento} {\bf 18} 393-400

\bibitem{Dirac 32 a}
Dirac PAM (1932) {\it The Principles of Quantum Mechanics} (Cambridge, at the Clarendon Press)

\bibitem{Gorringe 88 a}
Gorringe VM \& Leach PGL (1988) Lie point symmetries for systems
of second order linear ordinary differential equations {\it Qu\ae
stiones Mathematic\ae} {\bf 11} 95-117

\bibitem {Hamilton}
Hamilton WR (1834) On a general method in dynamics {\it
Philosophical Transactions of the Royal Society, Part II} 247-308.

\bibitem {Kanai 49 a}
Kanai E (1948) On the quantization of dissipative systems {\it Progress in Theoretical Physics} {\bf 3} 440-442

\bibitem{Lemmer 99 a}
Lemmer RL \& Leach PGL (1999/1420) A classical viewpoint on quantum chaos {\it Arab Journal of Mathematical Sciences} {\bf 5} 1-17

\bibitem {Mannheim 05 a}
Mannheim Phillip D \& Davidson Aharon (2005) Dirac quantization of the Pais-Uhlenbeck fourth order oscillator {\it Physical Review A} {\bf 71} 042110

\bibitem {Mannheim 07 a}
Mannheim PD (2007) Solution to the Ghost Problem in Fourth Order Derivative Theories {\it Foundations of Physics} {\bf 37} 532-571

\bibitem {Nucci}
Nucci MC (1996) The complete Kepler group can be derived by Lie
group analysis {\it J. Math. Phys.} {\bf 37} 1772-1775

\bibitem {Nucci again}
Gradassi A \&  Nucci MC (2007) Hidden linearity in systems for
competition with evolution in ecology and finance {\it J. Math.
Anal. Appl.} {\bf 333} 274-294

\bibitem {Nucci again again}
Nucci MC \& Leach PGL (2007) Lagrangians galore {\it J. Math.
Phys.} {\bf  48} 123510

\bibitem {Pais-Uhlenbeck 50 a}
Pais A \& Uhlenbeck GE (1950) On Field Theories with Non-Localized
Action {\it Physical Review} {\bf 79} 145-165

\bibitem {Schwartz 50 a}
Schwartz Laurent (1950) {\it Th\'eorie des Distributions} (Publications de l'Institut de Math\'ematiques de l'Universit\'e de Strasbourg, No 9) Vol 1

\bibitem {Schwartz 51 a}
Schwartz Laurent (1951) {\it Th\'eorie des Distributions} (Publications de l'Institut de Math\'ematiques de l'Universit\'e de Strasbourg, No 10) Vol 2

\bibitem {Temple}
Temple G (1955)  The Theory of Generalized Functions {\it
Proceedings of the Royal Society of London. Series A, Mathematical
and Physical Sciences} {\bf 228} 175-190

\bibitem {Whittaker 44 a}
Whittaker ET (1944) {\it A Treatise on the Analytical Dynamics of Articles and Rigid Bodies} (Dover, New York)

\end{thebibliography}
\end{document}